\documentclass{article}
\usepackage[T1]{fontenc} 
\usepackage[utf8]{inputenc} 
\usepackage{ismir,amsmath,amssymb,cite,url}
\usepackage{graphicx}
\usepackage{color}
\usepackage{algorithm}
\usepackage{algorithmicx}
\usepackage{caption}

\title{Towards Interpretable Polyphonic Transcription \\ with Invertible Neural Networks}

\oneauthor
    {Rainer Kelz$^1$, Gerhard Widmer$^{1,2}$ }
    {Austrian Research Institute for Artificial Intelligence (OFAI), Austria \\
      Institute of Computational Perception, Johannes Kepler University Linz, Austria \\ {\tt rainer.kelz@ofai.at}}
\def\authorname{Rainer Kelz, Gerhard Widmer}

\begin{document}
\maketitle
\begin{abstract}
We explore a novel way of conceptualising the task of polyphonic music transcription, using so-called invertible neural networks. Invertible models unify both discriminative and generative aspects in one function, sharing one set of parameters. Introducing invertibility enables the practitioner to directly inspect what the discriminative model has learned, and exactly determine which inputs lead to which outputs. For the task of transcribing polyphonic audio into symbolic form, these models may be especially useful as they allow us to observe, for instance, to what extent the concept of single notes could be learned from a corpus of polyphonic music alone (which has been identified as a serious problem in recent research). This is an entirely new approach to audio transcription, which first of all necessitates some groundwork. In this paper, we begin by looking at the simplest possible invertible transcription model, and then thoroughly investigate its properties. Finally, we will take first steps towards a more sophisticated and capable version. We use the task of piano transcription, and specifically the MAPS dataset, as a basis for these investigations.
  
\end{abstract}
\section{Introduction} \label{sec:introduction}

\begin{figure}
 \centerline{
   \includegraphics[width=1.1\columnwidth]{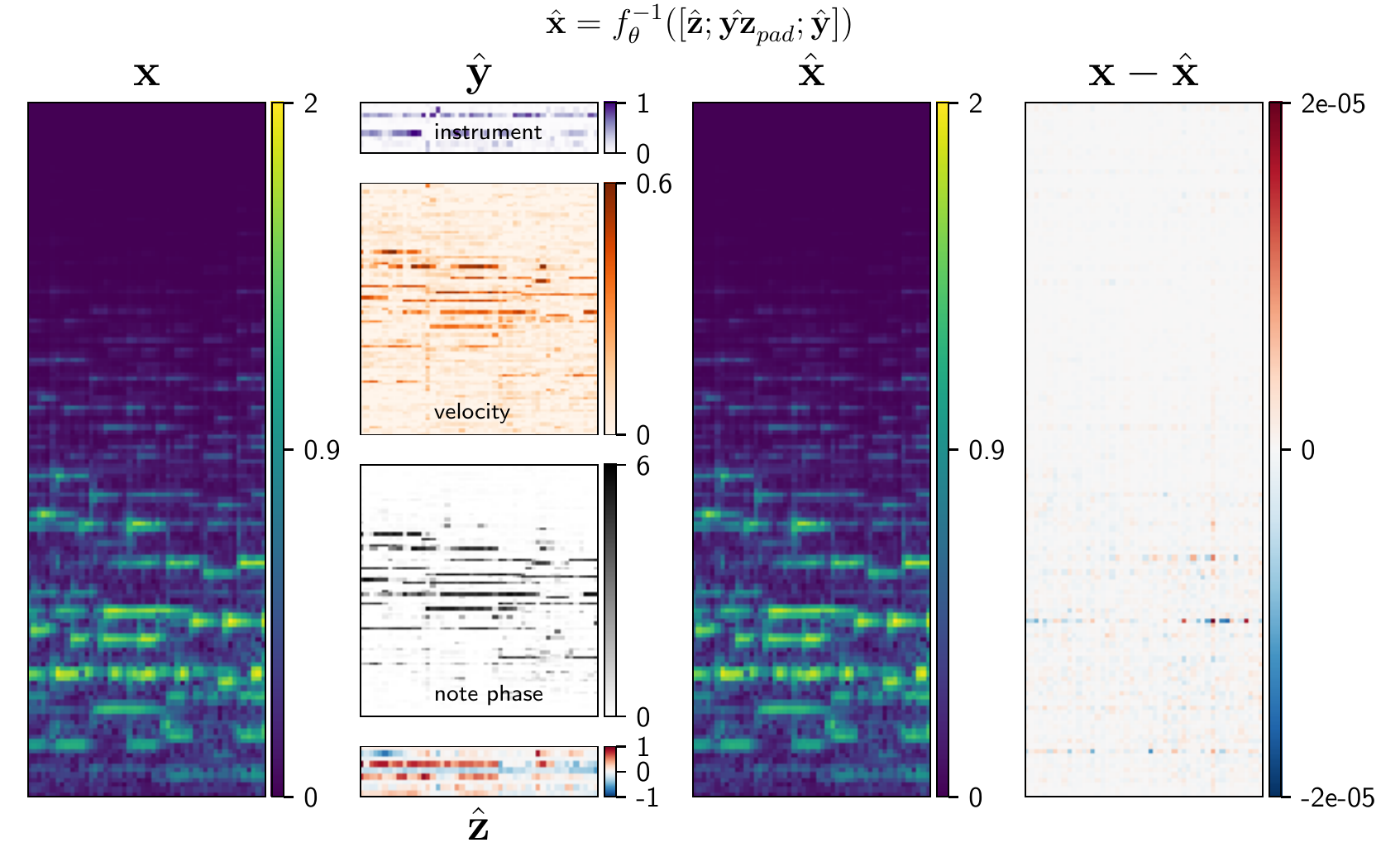}}

 \caption{Computing the framewise transcription $\hat{\mathbf{y}}$ and nuisance variables $\hat{\mathbf{z}}$ from spectrogram input $\mathbf{x}$. The predictions $[\hat{\mathbf{y}}; \hat{\mathbf{yz}}_{pad}; \hat{\mathbf{z}}]$ are then used to exactly reproduce $\hat{\mathbf{x}}$. The elementwise difference $\mathbf{x} - \hat{\mathbf{x}}$ is negligible. An in-depth discussion of this figure is deferred until \mbox{Section \ref{sec:ex_interpretability}}.}
 \label{fig:inn_z_hat_pad_y_hat_teaser}
\end{figure}

For practitioners who apply deep neural network models to music information retrieval tasks, interpretability of predictions is of great interest. Knowing what the model was able to learn from the data, and examining the underlying causes for a prediction increases trust in the model. Being aware of the reasons for a classification result allows us to discover whether the model has learned rules that would pass a basic sanity check with a domain expert, or if it has picked up on seemingly irrelevant factors present in the data, which made it possible for the network to solve the task in a different, unexpected, possibly unwanted way \cite{sturm_2014}. There are quite a few ways to obtain an explanation from a neural network. Several methods use the gradient of an output with respect to the input as a starting point, such as \cite{selvaraju_2017, sundararajan_2017}. There are also methods that aim to provide model agnostic explanations, for instance \cite{ribeiro_2016, plumb_2018}, and a specialization of one of the aforementioned methods to MIR systems \cite{mishra_2017} in particular.

Beyond providing explanations for predictions, a model should ideally be able to provide an answer to the question ``What do you consider representative examples for a concept of interest?''. Taking first steps towards producing models that are able to derive semantic information from the input, \textit{and} are able to answer this question, we explore invertible neural networks (INNs) with respect to interpretability of predictions, their potential to identify biases and confounding factors inherent in the training dataset, and ability to generate samples for a semantic concept of interest.

Additionally, we consider ways in which these models could enable us to locate ambiguous or uncertain predictions on unlabeled data, to provide eventual users of the MIR model with hints on where manual postprocessing of the predictions might be advisable. We choose to conduct our investigation in the context of polyphonic piano transcription and provide a first glimpse at the capabilities of INNs in Figure \ref{fig:inn_z_hat_pad_y_hat_teaser}. The input $\mathbf{x}$ to the INN is a magnitude spectrogram of an excerpt from a polyphonic piano piece, the output is split into a semantic part $\mathbf{y}$ containing variables of interest, and a nuisance part $\mathbf{z}$, optimistically containing all other factors of variation that are irrelevant for the MIR task the model was trained for. From these two output vectors, a hypothetical, perfectly converged invertible model can faithfully reproduce the input, down to a negligible numerical difference.

Invertible neural networks are parametrized, nonlinear and bijective functions, trainable from matched pairs, similar to any other neural network in a supervised learning task. The architectures we consider here are all constructed in such a way that the inverse is available in closed form. 

Networks designed in this fashion have a few desirable properties. They are both discriminative and generative models unified in one function, sharing one set of parameters. To put this into context, training a transcription system also yields a synthesizer, and vice versa training a synthesizer yields a transcription system. The term ``synthesizer'' is used rather loosely here, as the transcription system is trained with magnitude spectrograms.

This setup enables a direct interpretation of predictions by looking at what samples the model produces, conditioned on the predictions. This can potentially be extended until after eventual postprocessing steps, to see whether the generated samples are still close to the input in data space.

Furthermore, in order for a practitioner to understand whether the discriminatively trained network has learned to distinguish multiple concepts reasonably well, she can directly obtain samples from the model for each different concept. As an illustrative example, we choose the task of transcribing polyphonic audio into a symbolic format. This is a multi-label problem, assigning multiple note labels to each (quantized) point in time. Transcription systems based on neural networks are commonly learned from corpora containing large amounts of polyphonic music. Due to having the inverse available to us in closed form, we are able to sample all different single notes from the network to directly see whether the concept of single, isolated notes could be learned by training on our polyphonic corpus, or if multiple notes have been ``smeared'' together, and could not be disentangled from each other, or if the concept could not be learned at all. To the best of our knowledge, this is still an open problem that mostly affects polyphonic transcription systems based on neural networks, as discussed in \cite{kelz_2017}.

\section{Related Work} \label{sec:related_work}
Invertible neural networks were first introduced in \cite{deco_1995} and rediscovered in \cite{baird_2005}. They define a nonlinear, bijective mapping between inputs and outputs. They can be used to transform arbitrarily complex distributions into simple, factorized distributions. This concept became more widely known as normalizing flows, introduced in \cite{tabak_2010}, generalized in \cite{tabak_2013}, and has been used in \cite{dinh_2014} for density estimation, and improving variational inference in \cite{rezende_2015}. Various types of (more expressive) normalizing flows have been introduced in \cite{tomczak_2016, dinh_2017, berg_2018}. In \cite{kingma_2018} normalizing flows are employed as generative models for high resolution samples comparable to those produced by high resolution generative adversarial networks (GANs), e.g., \cite{karras_2018}.

With a greater focus on the invertibility aspect, \cite{ardizzone_2018} uses bijective architectures to approximate physical processes with a well defined forward model, in order to obtain the posterior distribution over inputs conditioned on desired outputs. We adopt parts of their terminology and training procedure. The differences will be discussed in more detail in Section \ref{sec:method}. In \cite{jacobsen_2018} injective and bijective i-RevNets are introduced, architectures similar to ResNets \cite{he_2016}, which are invertible up to the last layer. In \cite{jacobsen_2019}, fully invertible RevNets in conjunction with a new objective function are used to train classifiers which are more robust against adversarial attacks. We borrow their term ``nuisance'' variables to describe what information is supposed to end up in the output vector $\mathbf{z}$.

Distribution matching in this work is done using the sliced Wasserstein distance. Introduced in \cite{rabin_2012, bonneel_2015} as a distance measure for texture synthesis in a computer graphics setting, it has been used for encouraging the codes of autoencoders to follow a proposal distribution \cite{kolouri_2018}, and has also been directly applied to generative modeling of images, replacing the domain regressor in GANs \cite{deshpande_2018}.

Finally, we draw inspiration from \cite{maestro_2018} where a transcription-resynthesis system was introduced, consisting of three separately trained parts, a transcription system, a language model and a (neural) synthesizer.

\section{Method} \label{sec:method}
We adhere closely to the invertible neural network architectures described in \cite{ardizzone_2018}, with a minor modification to the training procedure that will be outlined after the formal introduction of invertible neural networks. Our notation also loosely follows the one used in \cite{ardizzone_2018}. Given a data space $\mathcal{X}$, a label space $\mathcal{Y}$ and a nuisance space $\mathcal{Z}$, we consider a directly invertible neural network as a parametrized function \mbox{$f_{\theta}: \mathcal{X} \rightarrow \mathcal{Y} \times \mathcal{Z}$} where we have access to its closed form inverse \mbox{$f_{\theta}^{-1}: \mathcal{Y} \times \mathcal{Z} \rightarrow \mathcal{X}$}.

The function $f_{\theta}$ maps the input into a label space that carries semantic information that we are interested in, and maps the rest of information into a nuisance space. Given both the semantic and nuisance information, we are able to obtain the input again via $f_{\theta}^{-1}$.

There are a few different ways such a function can be implemented in practice, and they all come with various different architectural constraints. We will first define a small invertible building block with relatively weak capabilities. These building blocks are then used to construct a more expressive function.

\begin{figure}
 \centerline{
 \includegraphics[width=\columnwidth]{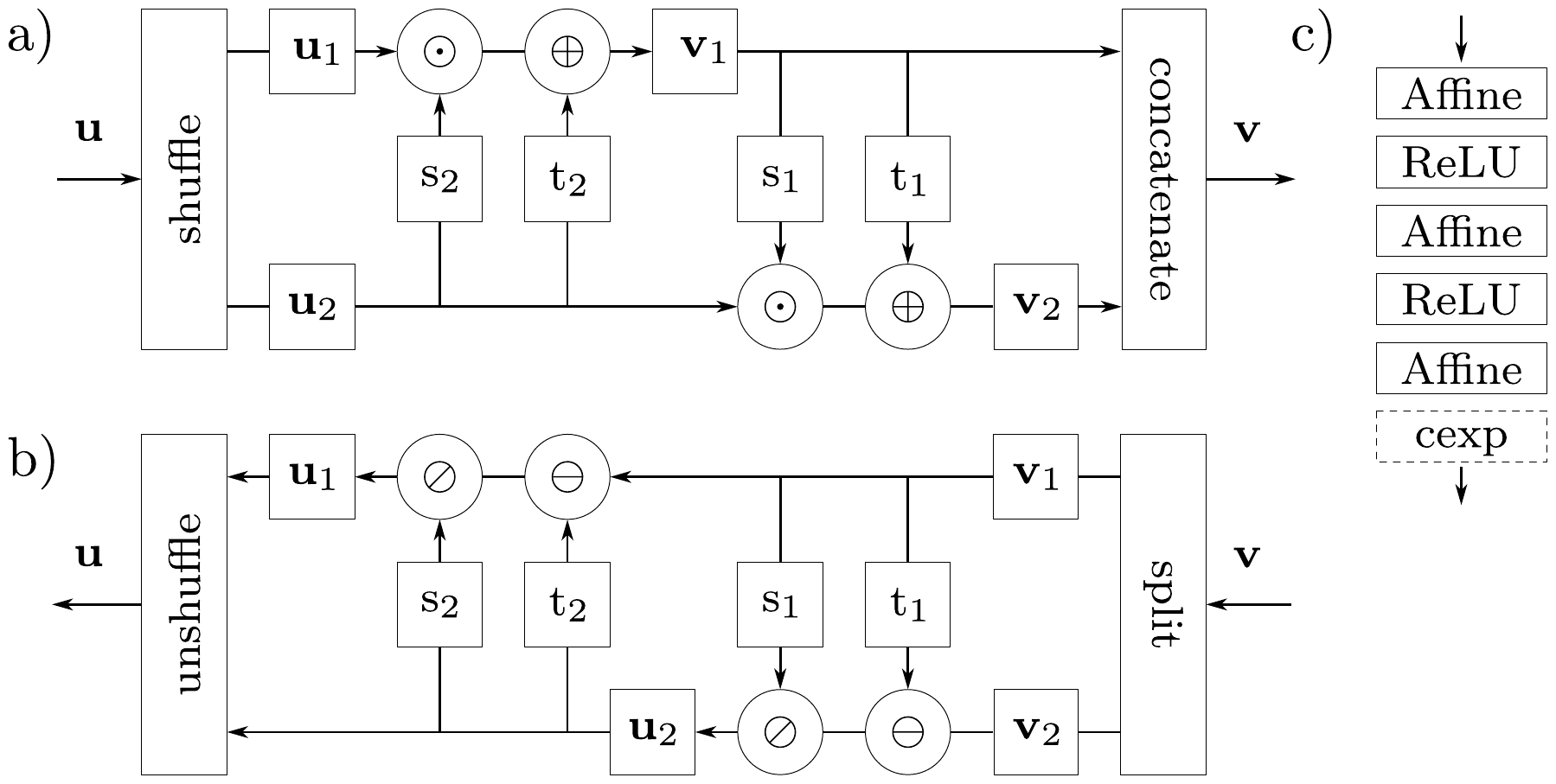}}
 \caption{This sketch depicts the structure of the particular version of affine coupling layers we use. \textbf{a}) The operations as they are applied in the forward direction. \textbf{b}) The operations as they are applied in the backward direction. \textbf{c}) The parametrization of the $\mathrm{s}_{1|2}$ and $\mathrm{t_{1|2}}$ transforms. The $\mathrm{cexp}$ function is only applied after the $\mathrm{s}_{1|2}$ transforms.}
 \label{fig:affine_coupling_layer}
\end{figure}

A necessary structural restriction to a bijective block is that the dimensionality of the input must match the dimensionality of the output. Another restriction concerns the inner workings of blocks, so their inverse is available in closed form. We adopt the affine coupling layer design in \cite{ardizzone_2018}, which is a more expressive version of the one in \cite{dinh_2017}. Its internal structure can be seen in Figure \ref{fig:affine_coupling_layer}. The layer takes as input a vector $\mathbf{u}$, whose dimensions are first shuffled with a fixed random permutation matrix and then split into two halves $\mathbf{u}_1, \mathbf{u}_2$. Dimension shuffling causes the splits to be different from layer to layer and facilitates interaction between components whose indices might be far apart in the input vector. The permutation matrix is inverted by simply transposing it.

Different operations are then applied to each half, after which the halves are concatenated again to yield the \mbox{output $\mathbf{v}$}. Equations \eqref{eq:affine_coupling_forward}~--~\eqref{eq:affine_coupling_backward} show the exact expressions used to compute results in both directions.

\begin{align} \label{eq:affine_coupling_forward}
  \mathbf{v}_1 & = \mathrm{cexp}(\mathrm{s}_2(\mathbf{u}_2)) \odot \mathbf{u}_1 \oplus \mathrm{t}_2(\mathbf{u}_2) \\
  \mathbf{v}_2 & = \mathrm{cexp}(\mathrm{s}_1(\mathbf{v}_1)) \odot \mathbf{u}_2 \oplus \mathrm{t}_1(\mathbf{v}_1) \\
  \mathbf{u}_2 & = (\mathbf{v}_2 \ominus \mathrm{t}_1(\mathbf{v}_1)) \oslash \mathrm{cexp}(\mathrm{s}_1(\mathbf{v}_1)) \\
  \mathbf{u}_1 & = (\mathbf{v}_1 \ominus \mathrm{t}_2(\mathbf{u}_2)) \oslash \mathrm{cexp}(\mathrm{s}_2(\mathbf{u}_2)) \label{eq:affine_coupling_backward}
\end{align}

Operations $\oplus, \ominus, \odot, \oslash$ (addition, subtraction, multiplication, division) are applied elementwise. The function $\mathrm{cexp}$ is defined as $\mathrm{cexp}(x) = \mathrm{exp}(c \cdot \mathrm{atan}(x))$ with $c > 0$ being a hyperparameter. Its purpose is to constrain the output to a reasonable range, and to prevent runaway growth of activations. The transforms $\mathrm{s}_{1|2}$ and $\mathrm{t}_{1|2}$ are arbitrarily parametrizable functions, modeling input dependent scaling and translation respectively. All transforms are implemented as standard neural networks, and are not required to be invertible, because the transformed half of the output vector can be inverted using the untransformed half. The network structures we use are shown in Figure \ref{fig:affine_coupling_layer}c. 

If the dimensionalities of input and output vectors do not match, the vectors are padded with zeros during inference, or small scale Gaussian noise during training, to encourage the network to ignore the additional padding dimensions, as done in \cite{ardizzone_2018}.

Each update of the model involves three passes, one forward pass, and two backward passes. Each pass has its own set of objective functions. The joint objective function to be minimized consists of a weighted sum of these terms. We specify the following notation: vectors are in boldface, writing vectors in square brackets separated by semicolons $[\mathbf{a}; \mathbf{b}]$ denotes concatenation. The vector $\mathbf{x}$ is the input to the model, $\mathbf{y}$ is the semantic part of the groundtruth, and $\mathbf{z}$ is a sample from a proposal distribution, which we choose to be $\mathcal{N}(\mathbf{0}, \mathbf{I})$. The padding vectors used during training are denoted as $\mathbf{x}_{pad}$ and $\mathbf{yz}_{pad}$ respectively, and are drawn from $\mathcal{N}(\mathbf{0}, \varepsilon)$ for each update, with $\varepsilon > 0$ a hyperparameter.  Symbols with a circumflex always refer to model outputs with a direct counterpart in the groundtruth. We denote a zero vector of a size appropriate in the context it appears in as $\mathbf{0}$. A sample from the model will be written as $\mathbf{x}_{sam}$. Equation \eqref{eq:full_inn_formally_forward} fully specifies all inputs and outputs for an invertible neural network used in the forward direction, equation \eqref{eq:full_inn_formally_backward} does the same in the backward direction, and \eqref{eq:full_inn_formally_sample} specifies how samples are drawn.

\begin{align}
  [\hat{\mathbf{z}}; \hat{\mathbf{yz}}_{pad}; \hat{\mathbf{y}}] & = f([\mathbf{x}; \mathbf{x}_{pad}]) \label{eq:full_inn_formally_forward} \\
  [\hat{\mathbf{x}}; \hat{\mathbf{x}}_{pad}] & = f^{-1}[\hat{\mathbf{z}}; \mathbf{yz}_{pad}; \hat{\mathbf{y}}] \label{eq:full_inn_formally_backward} \\
  [\mathbf{x}_{sam}; \hat{\mathbf{x}}_{pad}] & = f^{-1}[\mathbf{z}; \mathbf{0}; \mathbf{y}] \label{eq:full_inn_formally_sample}
\end{align}

\begin{algorithm}[t]
  \begin{algorithmic}
    \State \textbf{Let} $S \leftarrow 0$ and $\mathbf{A}, \mathbf{B} \in \mathbb{R}^{n \times d}$ (two samples)
    \State \textbf{For} $1\,..\,m$ \textbf{do}
    \State \ \ \ $\mathbf{p} \leftarrow \mathbf{p}' / \|\mathbf{p}' \|$ such that $\mathbf{p}' \sim \mathcal{N}(\mathbf{0}, \mathbf{I})$ and $\mathbf{p}' \in \mathbb{R}^{d \times 1}$
    \State \ \ \ $\textbf{a} \leftarrow \mathrm{sort}[\mathbf{A}\mathbf{p}]$; $\textbf{b} \leftarrow \mathrm{sort}[\mathbf{B}\mathbf{p}]$
    \State \ \ \ $S \leftarrow S + \|\mathbf{a} - \mathbf{b} \|_2^2 / n$
    \State \textbf{Return} $S/m$
  \end{algorithmic}
  \caption{Sliced Wasserstein Distance $d_{SWD}(\mathbf{A}, \mathbf{B})$}
  \label{alg:sliced_wasserstein_distance}
\end{algorithm}

Having defined these quantities, we can now proceed with defining the individual loss terms that will make up the joint objective function. Mean squared error \eqref{eq:loss_fit} is used to fit the labels from the groundtruth, and the reconstruction of the input \eqref{eq:loss_x_hat}. We deviate from \cite{ardizzone_2018} and use the sliced Wasserstein distance ($d_{SWD}$) \cite{rabin_2012} instead of the maximum mean discrepancy ($d_{MMD}$), to measure the distance between distributions, as we found it to be better behaved for high dimensional data. The intuition behind $d_{SWD}$ is to decompose the high dimensional optimal transport problem into $m$ 1-dimensional ones, by randomly projecting samples $\mathbf{A}$ and $\mathbf{B}$ onto lines, allowing the resulting 1-dimensional problems to be solved by computing the distance between sorted components. In equation \eqref{eq:loss_latent}, $d_{SWD}$ is used to minimize the distance between samples from the joint distribution over the outputs $[\hat{\mathbf{y}}; \hat{\mathbf{z}}]$ and samples from the joint distribution over the labels and the proposal distribution $[\mathbf{y}; \mathbf{z}]$. Please note that following the advice laid out in \cite{ardizzone_2018}, no gradient information from this objective is propagated back over $\hat{\mathbf{y}}$, to not unduly disturb the label fitting process. Informally stated, the purpose of including $\mathbf{y}$ and $\hat{\mathbf{y}}$ in the distribution matching process is to ``group'' samples together for which $\hat{\mathbf{z}}$ needs to follow a Gaussian distribution, resulting in the distributions $p(\hat{\mathbf{y}})$ and $p(\hat{\mathbf{z}})$ to gradually decouple, and become independent of each other, with the side effect of erasing label information from $\hat{\mathbf{z}}$. $d_{SWD}$ is also used in \eqref{eq:loss_backward}, to minimize the distance between the distribution of samples generated from the model and the groundtruth.

\begin{align}  
  \mathcal{L}_{\mathbf{y}}(\mathbf{y}, \hat{\mathbf{y}}) & = \mathrm{MSE}(\mathbf{y}, \hat{\mathbf{y}}) \label{eq:loss_fit} \\
  \mathcal{L}_{\hat{\mathbf{x}}}(\mathbf{x}, \hat{\mathbf{x}}) & = \mathrm{MSE}(\mathbf{x}, \hat{\mathbf{x}}) \label{eq:loss_x_hat} \\
  \mathcal{L}_{\mathbf{yz}}([\mathbf{y}; \mathbf{z}], [\hat{\mathbf{y}}; \hat{\mathbf{z}}]) & = \mathrm{SWD}([\mathbf{y}; \mathbf{z}], [\hat{\mathbf{y}}; \hat{\mathbf{z}}]) \label{eq:loss_latent} \\
  \mathcal{L}_{\mathbf{x}_{sam}}(\mathbf{x}, \mathbf{x}_{sam}) & = \mathrm{SWD}(\mathbf{x}, \mathbf{x}_{sam}) \label{eq:loss_backward} \\
  \mathcal{L}_{\mathbf{x}_{pad}}(\mathbf{x}_{pad}, \hat{\mathbf{x}}_{pad}) & = \mathrm{MSE}(\mathbf{x}_{pad}, \hat{\mathbf{x}}_{pad}) \label{eq:loss_x_pad} \\
  \mathcal{L}_{\mathbf{yz}_{pad}}(\mathbf{yz}_{pad}, \hat{\mathbf{yz}}_{pad}) & = \mathrm{MSE}(\mathbf{yz}_{pad}, \hat{\mathbf{yz}}_{pad}) \label{eq:loss_yz_pad}
\end{align}

Finally, the padding dimensions are taken care of with mean squared error terms \eqref{eq:loss_x_pad} and \eqref{eq:loss_yz_pad}, to encourage the network to disregard information in these dimensions. Following advice in \cite{ardizzone_2018}, the individual loss terms that sum up to the joint objective are weighted such that their magnitudes are approximately equal to each other, by restarting the optimization process multiple times and adjusting the weights until this condition is met.

\section{Experiments} \label{sec:experiments}
This section is split into multiple parts, starting out with a description of the data preparation procedure, followed by an empirical assessment of the usability of INNs for practitioners in \mbox{subsection \ref{sec:ex_usability}}, a critical examination of the interpretability of a trained model in \mbox{subsection \ref{sec:ex_interpretability}}, and finally an analysis of how well the concept of single notes could be learned from a polyphonic corpus in \mbox{subsection \ref{sec:ex_concept_understanding}}.

All model training, testing and generative sampling experiments were carried out with the MUS subset of the MAPS corpus \cite{emiya_2010}. This subset contains $210$ polyphonic piano pieces rendered with $7$ sample based synthesizers, and $60$ recordings of a YAMAHA Disklavier in two different recording conditions. After removing all synthesized pieces that also occur in the set of recordings, we are left with $139$ pieces for training, and the $60$ Disklavier recordings for testing, according to the procedure outlined in \cite{hawthorne_2018}. Evaluation measures are computed individually for each piece in the test set, and the mean over all pieces is reported. Groundtruth information is available as temporally aligned MIDI files. Sustain pedal control values are quantized, and the pedal considered fully engaged if its MIDI control value exceeds $64$. All offsets of notes that are sounding while sustain is in effect are extended in time, until the pedal is released again.

\begin{figure}
 \centerline{
   \includegraphics[width=1\columnwidth]{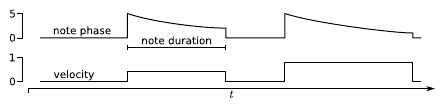}}
 \caption{This illustration shows how the note phase and velocity part of the label information $\mathbf{y}$ is derived for multiple notes played by a single key.}
 \label{fig:groundtruth}
\end{figure}

The label information $\mathbf{y}$ that the model has to learn is derived from the MIDI groundtruth and consists of $3$ parts: the note phase, its velocity and instrument information. For each piano key, the temporal evolution each note is modeled with an exponentially decaying curve, defined as \mbox{$\mathrm{curve}(\tau) = 0.99^{\tau} \cdot 5$}, with $0 \leq \tau < \mathrm{duration}$. It starts at the onset of a note, lasts for its duration, and drops off immediately after the offset. The velocity part is derived from the MIDI velocity value scaled into the interval $[0, 1]$. This procedure is also outlined in \mbox{Figure \ref{fig:groundtruth}}, and repeated for each of the $88$ piano keys. Finally, instruments are numbered from $0$ to $8$, corresponding to one of the $7$ sample banks or alternatively one of the two microphone conditions for the Disklavier recordings, and are one-hot encoded. For each (quantized) point in time $t$ all three parts are concatenated into the vector $\mathbf{y}_t$, having $9 + 88 + 88 = 185$ components.  This particular label vector derivation is chosen so that $\mathbf{y}_t$ contains all necessary information to generate spectrogram frames for different instruments and notes at the right volume and the right stage of a notes' temporal evolution without any additional context information from neighboring frames. The length of $\mathbf{z}_t$ was treated as a hyperparameter, and selected via cross validation on a small subset of the training set. Its length appears to have negligible influence given all the other settings, and was set to $9$ for all models subsequently used. The corresponding data $\mathbf{x}_t$ are magnitude spectrograms processed by a semi logarithmic filterbank, and the resulting bins $\mathbf{b}_t$ are elementwise processed by the function $\log(1 + \mathbf{b}_t)$, approximating human loudness perception to finally yield a vector $\mathbf{x}_t$ of length $144$. All spectral feature extraction and filtering is done with the \verb|madmom| library \cite{madmom}. The frame rate at which pairs $(\mathbf{x}_t, \mathbf{y}_t)$ are extracted from the audio and MIDI files is $25$ frames per second. As all input is processed in a framewise fashion everywhere, we omit the subscript $t$, denoting time in frames, for all plots and most equations to not add additional clutter. To increase the capacity of the INN, we add zero padding vectors to both input ($\mathbf{x}_{pad}$) and output dimensions ($\mathbf{yz}_{pad}$), so the number of components in the padded vectors sum up to $256$ in total. Training follows the procedure outlined in \mbox{Section \ref{sec:method}}, and all code is released\footnote{https://github.com/rainerkelz/ISMIR19} to facilitate reproducability.

\subsection{Usability for MIR tasks} \label{sec:ex_usability}
As a kind of quantitative viability test, the capability of INNs (in combination with simple temporal models) to produce predictions useful to MIR practitioners, is examined. We train small recurrent networks (RNNs) on the framewise predictions $\hat{\mathbf{y}}$ obtained from the INN, in the hope to obtain cleaner, denoised framewise predictions $\hat{\mathbf{y}}'$. The types of RNN cells we use are either LSTM \cite{hochreiter_1997} or GRU \cite{cho_2014} cells. In a first attempt, RNNs with very limited capacity - $4$ hidden units / cells for all keys - are employed. An input sequence to the RNN consists of the note phase and the velocity part of a \textit{single} key over the whole length of the piece, leading to an input dimension of $2$. The RNNs should output a smoothed, denoised version of the note phase and velocity sequence, and an additional framewise note activity indicator between $0$ and $1$, and thus all have $3$ outputs. The binary piano roll used to compute framewise performance measures is obtained by thresholding the note activity indicator output of the RNNs at $0.5$. Training proceeds one full sequence at a time, picked uniformly at random from all $139 \cdot 88$ single key sequences derived from all pieces in the training set. The models with highest $\mathrm{F}_1$-measure on a subset of the training set are then evaluated on the test set.

In order to evaluate framewise performance for a piece, we produce framewise transcriptions for all keys with the INN. Each key is then separately smoothed, denoised, and its activity is inferred over the length of the piece. We report the results for the small models in \mbox{Table \ref{tab:rnn_results}}, suffixed with ``(S)''. We would like to note that there was next to no hyperparameter tuning done, aside from getting the learn rates for the two different RNN cell types approximately in the right regime. A slightly larger version uses $3$ layers of bi-directional GRU cells with $8$ hidden units, and dropout \cite{srivastava_2014} with a probability of $0.5$, applied to the output of each recurrent layer, before it is passed on to the next. Results in the table for this type of RNN are suffixed with ``(L)''. The INN has $5$ invertible layers, and $990.720$ parameters in total. The parameter counts for the recurrent model variants \mbox{``GRU (S)''}, \mbox{``LSTM (S)''} and \mbox{``biGRU (L)''} are $111$, $143$ and $3123$ respectively.

We can see that the combination framewise \mbox{INN + biGRU} performs on par with the \mbox{CNN + RNN-NADE} combination in terms of framewise performance, and slightly outperforms the standalone CNN. The last three rows in \mbox{Table \ref{tab:rnn_results}} are taken from \cite{hawthorne_2018}, who re-implemented the approaches in \cite{kelz_2016} and \cite{sigtia_2016}, and ostensibly performed additional hyperparameter tuning to improve upon the original results. They also provide the current state of the art results for this train and test protocol in the last row, achieved by supplying an additional onset target to the network during training.

\begin{table}
 \begin{center}
 \begin{tabular}{|c|c|c|c|}
  \hline
  Method & $\mathrm{P}$ & $\mathrm{R}$ & $\mathrm{F}_1$ \\
  \hline
  INN + GRU (S)  & $79.74$ & $63.73$ & $70.84$ \\
  INN + LSTM (S)  & $80.12$ & $63.91$ & $71.10$ \\
  \hline
  INN + biGRU (L)  & $81.72$ & $64.81$ & $72.29$ \\
  \hline
  CNN only \cite{kelz_2016, hawthorne_2018} & $81.18$ & $65.07$ & $71.60$ \\
  CNN + RNN-NADE \cite{sigtia_2016, hawthorne_2018} & $71.99$ & $73.32$ & $72.22$ \\
  CNN + LSTM \cite{hawthorne_2018} & $88.53$ & $70.89$ & $78.30$  \\
  \hline
 \end{tabular}
\end{center}
 \caption{Framewise performance of different combinations of acoustic and temporal models on the testset.}
 \label{tab:rnn_results}
\end{table}

\subsection{Interpretability of results} \label{sec:ex_interpretability}
This section considers how the ability to modify the output of the model, and then using it in the backward direction, can assist the practitioner in determining the causes in the data that led to a particular prediction. We start with a thought experiment, and get closer to reality step by step. Let us assume the model works perfectly, and given an input $\mathbf{x}$, the model routes all semantic information (note phases, velocities, instrument) into the $\hat{\mathbf{y}}$ vector, all nuisance information (other acoustic variability, such as microphone characteristics, room reverberation or actual noise) ends up in $\hat{\mathbf{z}}$ which is distributed as $\mathcal{N}(\mathbf{0}, \mathbf{I})$, and the padding vector $\hat{\mathbf{yz}}$ is exactly zero. Sampling \mbox{$\mathbf{z}_s \sim \mathcal{N}(\mathbf{0}, \mathbf{I})$} and using $f^{-1}([\hat{\mathbf{y}}, \mathbf{0}, \mathbf{z}_s])$ to obtain the corresponding input $\mathbf{x}_s$ will change \textit{only nuisance characteristics} in the input. This would also mean that we have \textit{full control} over the semantic content of the input. We could add or delete notes \textit{in the input} simply by adding or zeroing them out in $\hat{\mathbf{y}}$, much like we can insert or delete a symbolic MIDI note, the implication being that every output has a \textit{directly} interpretable correspondence in the input.

A closer look at \mbox{Figure \ref{fig:inn_z_hat_pad_y_hat_teaser}} reveals what can realistically be achieved just by training a framewise INN with a rather limited amount of polyphonic data. It is immediately noticeable that $\hat{\mathbf{z}}$ does \textit{not} appear to be normally distributed. There are still patterns discernible, making it apparent that it still contains semantic information. Similar patterns also exist in $\hat{\mathbf{yz}}_{pad}$ (not shown).

\begin{figure}[t]
 \centerline{
   \includegraphics[width=1.1\columnwidth]{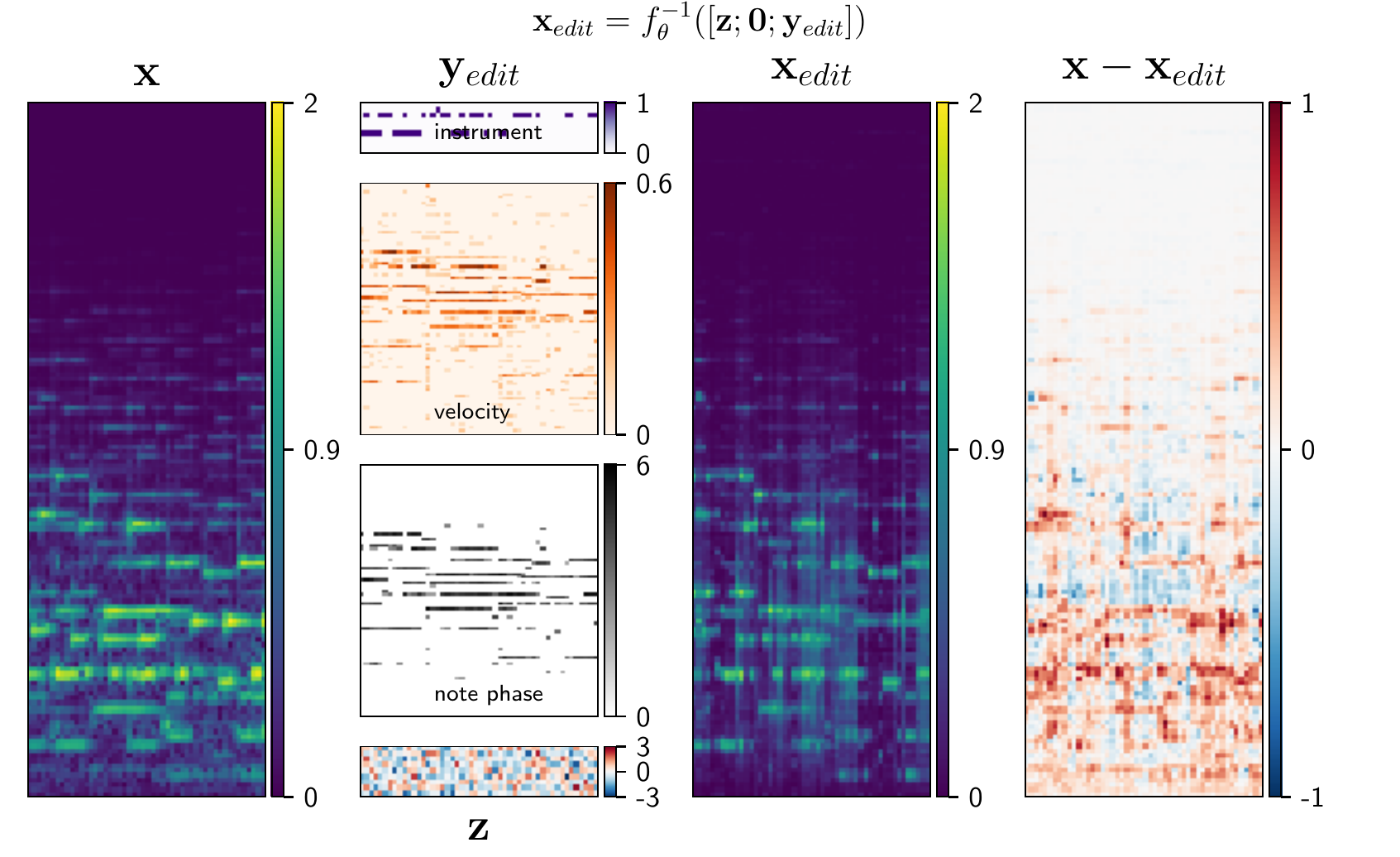}}
 \caption{Gradually denoising the predictions with simple, ad-hoc rules, zero padding and sampling $\mathbf{z} \sim \mathcal{N}(\mathbf{0}, \mathbf{I})$. In practice, this would be done iteratively, always keeping an eye on how the overall structure of the input is affected.}
 \label{fig:inn_z_zero_y_edit}
\end{figure}

\begin{figure}[t]
 \centerline{
   \includegraphics[width=1.1\columnwidth]{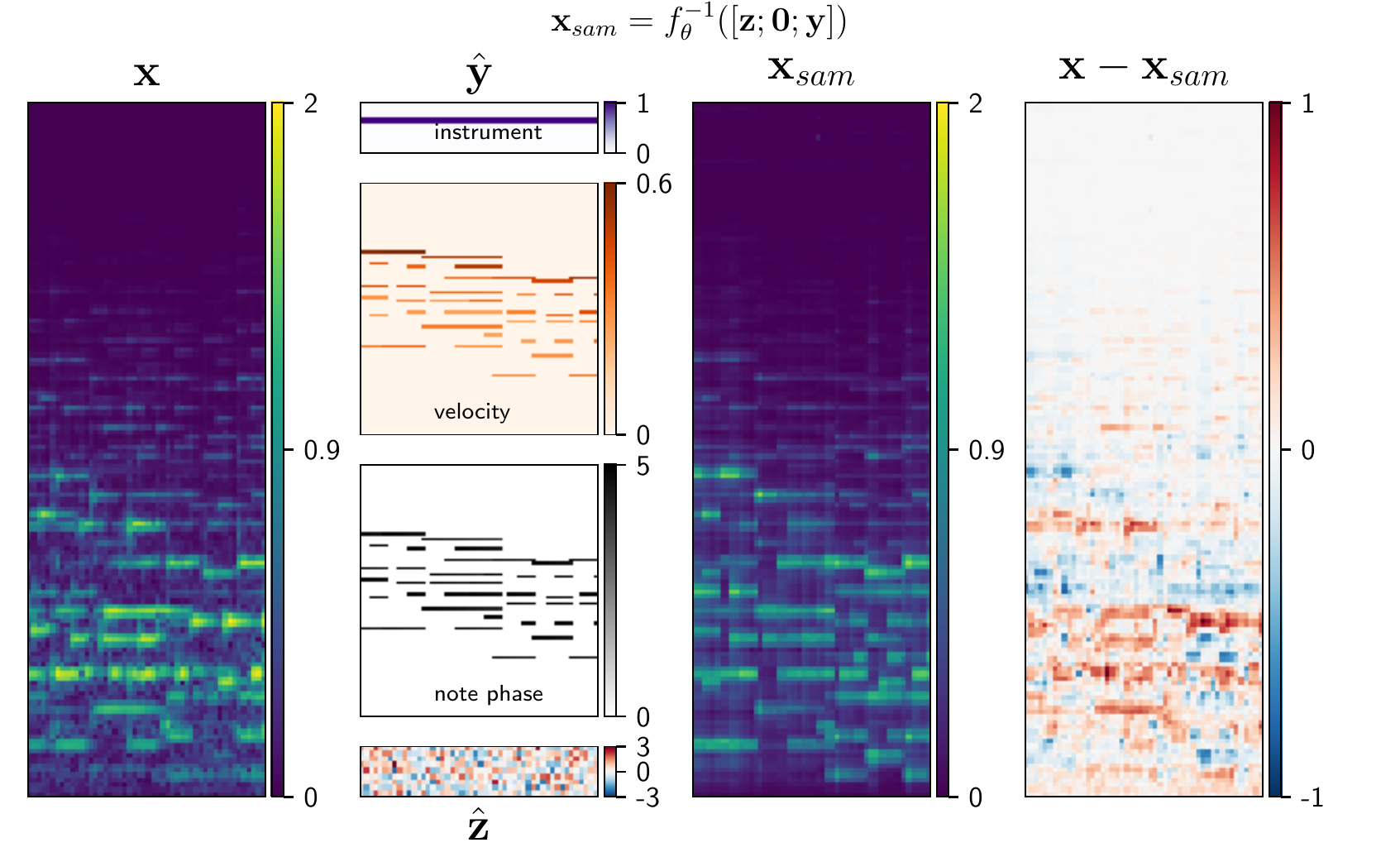}}
 \caption{The hypothetical case where all predictions could be perfectly denoised. For purely demonstrational purposes, this was accomplished through consultation of an oracle, but one could imagine this to be achievable through interaction with the system. Although quite a bit of detail is missing, the majority of the structure in the original input is nonetheless recognizable in the generated sample.}
 \label{fig:inn_z_samp_zero_y_hat}
\end{figure}

It is also observable that the information that is routed into $\hat{\mathbf{y}}$ is somewhat noisy. We can now attempt to ``separate the wheat from the chaff'' by using the INN in the backward direction with cleaned up predictions. \mbox{Figure \ref{fig:inn_z_zero_y_edit}} shows what happens when the predictions are partially denoised by setting all predictions below a certain threshold to zero, ignoring the padding vector by zeroing it out, and sampling $\mathbf{z}$ from a unit normal distribution. These simple, ad-hoc rules cannot get rid of all the noise and discontinuities in the predictions, but are useful to determine which outputs can be ignored by observing their (collective) impact on the sample. \mbox{Figure \ref{fig:inn_z_samp_zero_y_hat}} depicts what can be generated by the model, assuming that the denoising process of the predictions were perfect, by consulting an oracle about the true contents of $\mathbf{y}$. In all figures discussed in this section, the same excerpt from the test set was used, meaning the model has never seen any of the examples during training.

\subsection{Concept Understanding} \label{sec:ex_concept_understanding}
\begin{figure}
 \centerline{
   \includegraphics[width=1.1\columnwidth]{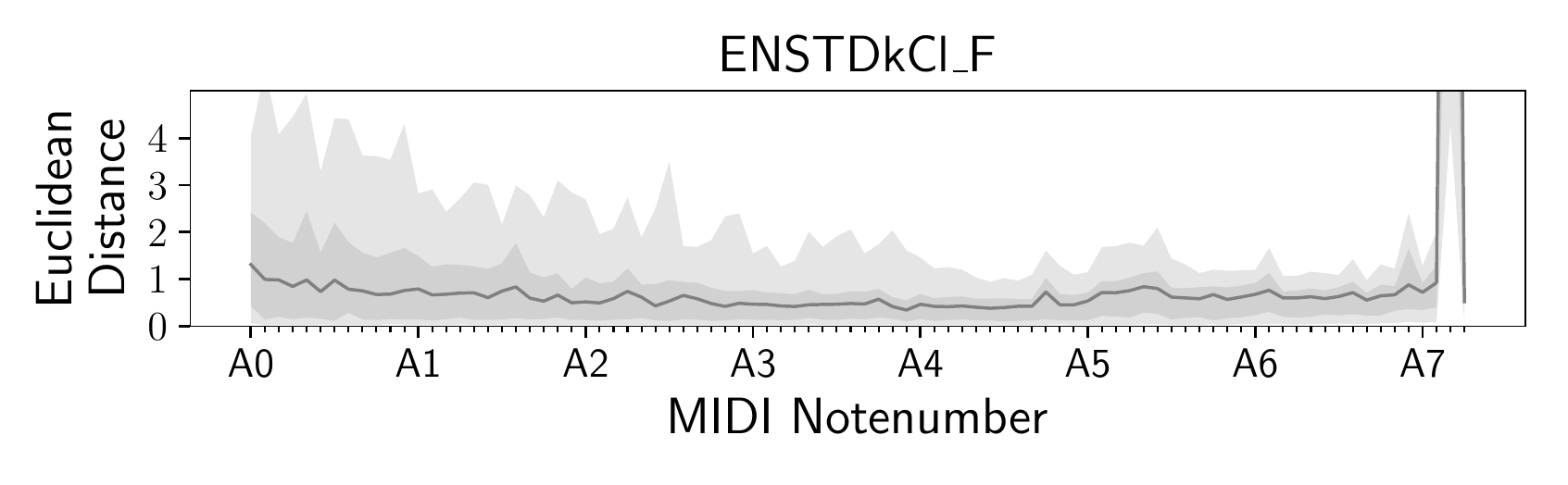}}
 \caption{Interquantile ranges of the framewise Euclidean distances between isolated reference notes and samples from the model. The reference notes stem from an unseen instrument.}
 \label{fig:concept_single_notes_test}
\end{figure}

\begin{figure}
 \centerline{
   \includegraphics[width=1.1\columnwidth]{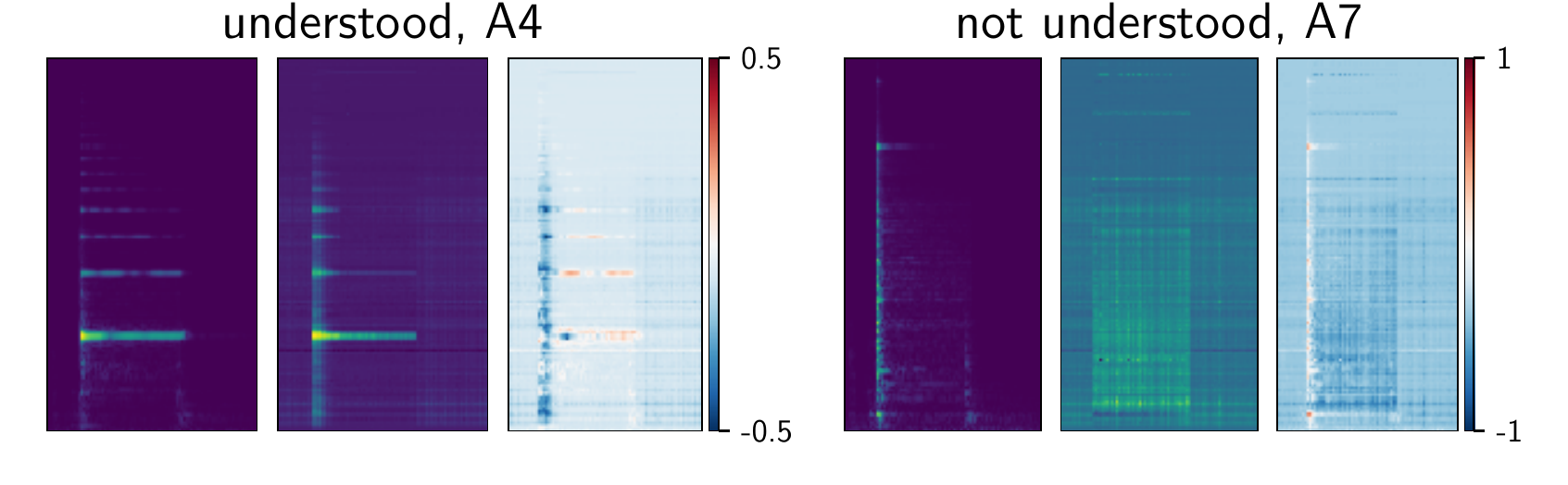}}
 \caption{Samples from the model for single notes.}
 \label{fig:concept_good_bad}
\end{figure}

Returning to a question raised in the introduction, in this section the model will be systematically queried about specific semantic concepts. Arguably, a polyphonic transcription system should be able to transcribe isolated notes. The MAPS dataset provides both renderings and recordings of isolated notes, which we utilize to formulate our queries. For each of the $88$ keys, $30$ samples are drawn from the model, using the groundtruth $\mathbf{y}$ paired with the reference recording for the key and $\mathbf{z} \sim \mathcal{N}(\mathbf{0}, \mathbf{I})$. This ensures that each sample has the same length as the reference. For each frame at time $t$ in a sample, the Euclidean distance to the corresponding frame of the reference recording is measured, and $p$-quantiles are computed on the resulting lists of framewise distances, with $p \in \{0.05, 0.25, 0.5, 0.75, 0.95\}$. \mbox{Figure \ref{fig:concept_single_notes_test}} depicts the interquantile range $[0.05, 0.95]$ as light gray, the range $[0.25, 0.75]$ in a darker shade, and the median as a black line. It becomes immediately apparent that samples for rarely occuring (possibly omitted) notes, such as those in the lower and higher octaves, are highly dissimilar from the reference recordings, and indicate that these particular isolated notes could not be learned by the network (\mbox{Figure \ref{fig:concept_good_bad}}). Admittedly, this question could have been answered for a regular feedforward network as well, but would have necessitated \textit{more labeled} reference data of the \textit{same} instrument. The ability to sample from the model allows us to sidestep the rather cumbersome way of aggregating prediction errors, as was necessary in \cite{kelz_2017}, to arrive at a similar conclusion.


\subsection{Improving Models with temporal context} \label{sec:ex_sophisticated}
The invertible models investigated so far all take single frames as input, without temporal context information. We trained fully invertible RevNets \cite{jacobsen_2018, jacobsen_2019} on a variety of different context lengths, but were not yet able to observe either quantitative or qualitative improvements over the framewise models. RevNets tend to become rather large in terms of the number of parameters, input and output padding is not as straightforward as for framewise models, the input and output space dimensionality is much larger, making the sliced Wasserstein distance gradually less effective due to an increase in necessary computational resources, which in turn further slows down training. Finally, the amount of training data we use may simply be insufficient for higher capacity models. However, we believe that all these issues have appropriate remedies. An immediate next step would be to apply the same models to the much larger \mbox{MAESTRO} dataset \cite{maestro_2018}. We leave these steps for future work though.

\section{Conclusion} \label{sec:conclusion}
The viability of invertible neural networks for a selected MIR task was shown quantitatively in terms of transcription performance and a brief numerical analysis of single concept understanding. A qualitative investigation of the direct interpretability of outputs back in input space was conducted. There is ample room for improvement, such as using an adversarial distance for distribution matching in both input and output space, or alternatively using the independent cross-entropy objective from \cite{jacobsen_2019} in latent space. The objective for the semantic part of the output space could be similarly augmented to encourage the disentanglement of (predictions for) individual notes. Another obvious improvement would be to skip the computation of filtered spectrograms altogether, and feed in waveforms to obtain models that can in turn generate waveforms we can directly listen to.

Beyond the interpretability aspect, we are confident that invertible neural networks will prove to be useful for other MIR tasks as well, such as musical content-aware style transfer (this is already doable with the models used in this work, by simply changing the instrument encoding when sampling, although changing from one piano to a different one is not as exciting as changing it into a trumpet). These models could also be adapted for (blind) source separation, to name only two examples.

\section{Acknowledgements}
We would like to express our deep gratitude to \mbox{Lynton} \mbox{Ardizzone} for pointing us in the right direction and \mbox{Hamid} \mbox{Eghbal-zadeh} for noting the pitch content-aware style transfer angle. This research has received funding from the European Research Council (ERC) under the European Union’s Horizon 2020 research and innovation programme (Grant agreement No. 670035, project \mbox{CON ESPRESSIONE}). The Tesla K40 used for this research was donated by the NVIDIA Corporation.

\bibliography{master}

\end{document}